\documentclass[twocolumn,showpacs,amsmath,amssymb,floatfix,prl,aps,superscriptaddress,showkeys]{revtex4}

\usepackage{graphicx}
\usepackage{epsfig}
\usepackage{dcolumn}
\usepackage{bm}
\usepackage{times}

\begin{document}
\title{Quantum bouncer with quadratic dissipation}

\author{Gabriel Gonz\'alez}\email{ggonzalez@physics.ucf.edu}
\address{NanoScience Technology Center, University of Central
Florida, Orlando, FL 32826, USA}
\address{Department of Physics, University of Central
Florida, Orlando, FL 32816-2385, USA}

\begin{abstract}
The energy loss due to a quadratic velocity dependent force on a quantum particle bouncing on a perfectly reflecting surface is obtained for a full cycle of motion. We approach this problem by means of a new effective phenomenological Hamiltonian which corresponds to the actual energy of the system and obtained the correction to the eigenvalues of the energy in first order quantum perturbation theory for the case of weak dissipation.  \\ \\
{\it Resumen} : La p\'erdida de energ{\'i}a debido a una fuerza proporcional al cuadrado de la velocidad se obtiene para el movimiento de una part{\'i}cula en el campo gravitacional uniforme. Se propone un nuevo Hamiltoniano efectivo para obtener las correciones a los eigenvalores utilizando la teor{\'i}a de perturbaciones para el caso de disipaci\'on d\'ebil.
\end{abstract}

\maketitle

\section{Introduction}
\label{sec:introduction}
Recently there has been a lot of attention to the experimental realization of gravitationally bound quantum systems \cite{obs,obs1}. Quantum states in the Earth's gravitational field have been observed with ultra-cold neutrons falling under gravity \cite{obs2}. The toy model known as the {\it quantum bouncer} which describes a quantum particle bouncing in a linear gravitational field has been used to compare the theoretical calculation with the experimental results. In all these experiments they have found energy losses due to different processes \cite{obs3}. One way of simulating this energy loss is by adding an additional external
velocity-dependent force acting on the conservative system
and transforming it into a non conservative system. The resulting classical dissipative
system thus contains this phenomenological
velocity-dependent force. 
The classical Hamiltonian is used as a basis for the so-called canonical quantization by obtaining the corresponding Hamiltonian operator.
For dissipative systems, i.e. systems where mechanical energy is lost due to frictional forces, difficulties arise in defining a Hamiltonian function \cite{GG,GG1,GG2}. Although formal Lagrangian functions yielding the correct equations of motion can always be given for one dimensional non conservative systems, one can not always find the corresponding Hamiltonian, and even if the corresponding Hamiltonian exists we find problems in their physical interpretation. This becomes even more obvious when these Hamiltonians are quantized in the usual canonical way \cite{GG3}.\\
In this article we propose a new effective phenomelogical Hamiltonian for the motion of a particle in a uniform gravitational field and under a frictional force which is proportional to the square of the velocity. This new Hamiltonian allows a physical interpretation in terms of the energy of the system. Using this Hamiltonian we obtain the energy loss for a full cycle of motion for the quantum bouncer with quadratic dissipation by means of canonical quantization and perturbation theory.

\section{Model Hamiltonian}
\label{sec:model}
Suppose we drop a particle of mass $m$ from a distance $d$ above the surface of the Earth and we consider that during its motion there is a frictional force which is proportional to the square of the particle's velocity. The equation of motion which describes the dynamics of the particle is given by 
\begin{equation}
m\frac{dv}{dt}=-mg-\gamma v |v|=
\left\{ \begin{array}{ll}
												-mg-\gamma v^2 & \mbox{if $v>0$}\\
												-mg+\gamma v^2 & \mbox{if $v<0$},
												\end{array} \right.
\label{eq1}
\end{equation} 
where $\gamma>0$ is the dissipation parameter. One can verify that the Hamiltonian for Eq. (\ref{eq1}) is given by
\begin{equation}
H_{\downarrow}=\frac{p^2}{2m}e^{2\gamma x/m}-\frac{m^2 g}{2\gamma}[e^{-2\gamma x/m}-1],
\label{eq2a}
\end{equation} 
\begin{equation}
H_{\uparrow}=\frac{p^2}{2m}e^{-2\gamma x/m}+\frac{m^2 g}{2\gamma}[e^{2\gamma x/m}-1],
\label{eq2b}
\end{equation}
where $H_{\downarrow}$ and $H_{\uparrow}$ represents the Hamiltonian for the case when $v<0$ and $v>0$ respectively.
The canonical quantization of $H_{\downarrow}$ and $H_{\uparrow}$ has been obtained and studied by several authors \cite{obs4, obs5, obs6, obs7}, but a number of well known difficulties arise when one tries to interpret the results using this type of Hamiltonian in the quantum regime \cite{obs8, obs9}, in particular we see that $H_{\downarrow}$ and $H_{\uparrow}$ can be obtained from each other by making the substitution $\gamma\rightarrow-\gamma$, due to this symmetry and for the case of weak dissipation, the correction to the eigenvalues for a full cycle cancels out for odd powers in the dissipation parameter and only even powers of the dissipation parameter remain by using quantum perturbation theory. \\
The main goal of this article is to find a new Hamiltonian that describes Eq. (\ref{eq1}) and calculate the change in energy for a full cycle for the case of weak dissipation using quantum perturbation theory. To do this we are going to use the fact that the dynamics of Eq. (\ref{eq1}) can be completely determined in the classical sense and allows us to express the square of the velocity in terms of the particle's position 
\begin{equation}
v^{2}_{\downarrow}(x)=\frac{mg(1-e^{-2\gamma(d-x)/m})}{\gamma}, 
\label{eq3a}
\end{equation}
\begin{equation}
v^{2}_{\uparrow}(x)=\frac{mg(2e^{-2\gamma x/m}-e^{-2\gamma(d+x)/m}-1)}{\gamma},
\label{eq3b}
\end{equation} 
where we have taken into account that $v(d)=0$ and that the particle undergoes a perfectly elastic collision when it bounces on the surface of the Earth. One can easily see that when the dissipation parameter goes to zero, i.e. $\gamma \rightarrow 0$, we obtain the usual kinematic expressions for the square of the velocity for a particle in a uniform gravitational field. \\ Plugging Eq. (\ref{eq3a}) and Eq. (\ref{eq3b}) into Eq. (\ref{eq1}) we have 
\begin{equation}
m\frac{dv_{\uparrow}}{dt}=-mg(2e^{-2\gamma x/m}-e^{-2\gamma(d+x)/m}),
\label{eq4a}
\end{equation}
\begin{equation}
m\frac{dv_{\downarrow}}{dt}=-mge^{-2\gamma(d-x)/m}.
\label{eq4b}
\end{equation}
In this way we can construct the Hamiltonian for Eq. (\ref{eq4a}) and Eq. (\ref{eq4b})	in the usual way which are given by
\begin{equation}
{\cal H}_{\downarrow}=\frac{p^2}{2m}+\frac{m^2 g}{2\gamma}e^{-2\gamma d/m}\left(e^{2\gamma x/m}-1\right),
\label{eq5a}
\end{equation}																		  
\begin{equation}
{\cal H}_{\uparrow}=\frac{p^2}{2m}+\frac{m^2 g}{2\gamma}\left(e^{-2\gamma d/m}-2\right)\left(e^{-2\gamma x/m}-1\right),
\label{eq5b}
\end{equation}
where we have taken into account that one must obtain the usual Hamiltonian when the dissipation parameter goes to zero. \\
Using Eq. (\ref{eq5a}) and Eq. (\ref{eq5b}) we can write an effective Hamiltonian in the following form
\begin{equation}
H_{eff}=\frac{{\cal H}_{\uparrow}+{\cal H}_{\downarrow}}{2}+\frac{p}{|p|}\left(\frac{{\cal H}_{\uparrow}-{\cal H}_{\downarrow}}{2}\right).
\label{EH}
\end{equation} 

\section{Canonical Quantization}
\label{sec:CQ}
\begin{figure}[ht]
\includegraphics[width=6.2cm]{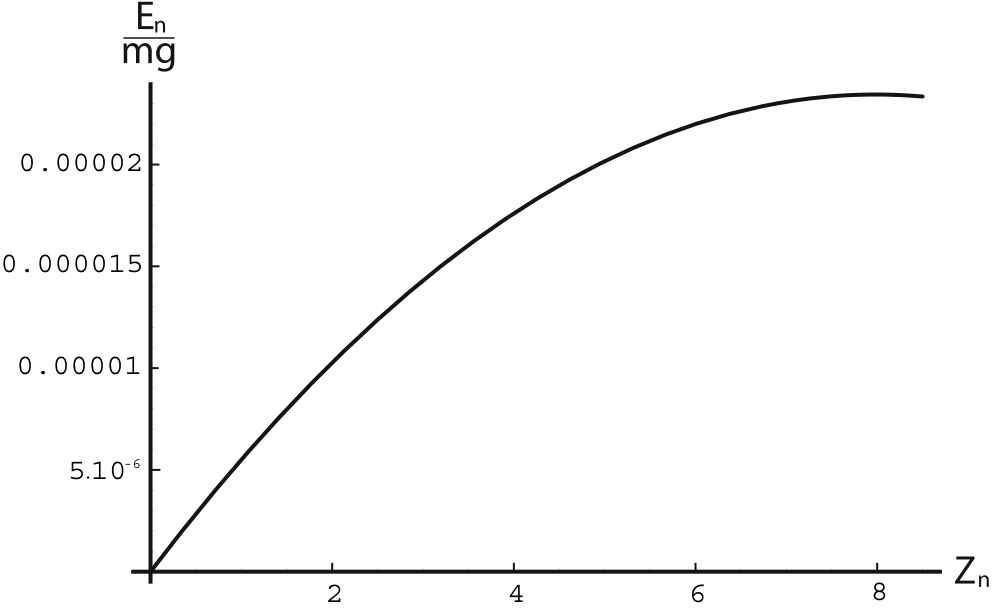}
\caption{The graph shows  the energy loss of a neutron in a gravitational field where $\ell_{g}$=5.57$\mu$m, $m$=1.674$\times 10^{-27}$kg, $\gamma\sim 10^{-23}$kg/m and $d$=3mm. When $\gamma=0$ we have for the two lowest quantum states $E_{n}^{(0)}/mg = (13.7,24.0)\mu m$ and $z_{n}=(2.34,4.09)$, respectively.}
\vspace{-1cm}
\label{Fig1}
\end{figure}

To see the effects of dissipation in the eigenvalues of the quantum bouncer we are going to consider the case when we have weak dissipation such that ${\cal H}_{\uparrow}\approx {\cal H}_{\downarrow}$, for this case we can neglect the second term of Eq. (\ref{EH}) since it cancels during a full cycle. Expanding Eq. (\ref{eq5a}) and Eq.~(\ref{eq5b}) in a Taylor series in the following way
\begin{equation}
{\cal H}_{\downarrow}=\frac{p^2}{2m}+A\left(x+\frac{2\gamma}{m}x^2+\cdots \right),
\label{eq6a}
\end{equation}																		  
\begin{equation}
{\cal H}_{\uparrow}=\frac{p^2}{2m}-B\left(-x+ \frac{2\gamma}{m}x^2- \cdots\right),
\label{eq6b}
\end{equation}
where $A=mge^{-2\gamma d/m}$ and $B=mg(2-e^{-2\gamma d/m})$, and keeping only the first two terms we end up with the following effective Hamiltonian 
\begin{equation}
H_{eff}=\frac{p^2}{2m}+\left(\frac{A+B}{2}\right)x+\left(\frac{A-B}{2}\right)\left(\frac{2\gamma x^2}{m}\right).
\label{EH1}
\end{equation}
Treating the last term in Eq. (\ref{EH1}) as a perturbation we can estimate the correction to
the energy due to dissipation by using quantum perturbation theory where the unperturbed Hamiltonian is given by 				\begin{equation}
\hat H_{0}=\frac{\hat{p}^2}{2m}+mg\hat{x}.
\label{eq7}
\end{equation}
It is very well known that the normalized eigenfunctions for Eq. (\ref{eq7}) are given by Airy functions and its first derivative evaluated at its $n$th zero \cite{obs10}
\begin{equation}
\psi_{n}^{(0)}(z)=\frac{Ai(z-z_{n})}{|Ai'(-z_{n})|},
\label{eq8}
\end{equation}
where $z=x/\ell_{g}$  and $z_{n}=(E_{n}^{(0)}/mg\ell_{g})$ are defined in terms of the {\it gravitational length} $\ell_{g}=(\hbar^2/2m^2 g)^{1/3}$, respectively. \\
Using the above results we can determine the correction to the eigenvalues of the energy as
\begin{equation}
\delta E_{n}^{(1)}=-2\gamma g\ell_{g}^2(1-e^{-2\gamma d/m})\langle \psi_{n}^{(0)}|z^2|\psi_{n}^{(0)}\rangle, 
\label{eq9}
\end{equation}
and using the fact that \cite{obs10}
\begin{equation}
\langle \psi_{n}^{(0)}|z^2|\psi_{n}^{(0)}\rangle=\frac{8}{15}z_{n}^2,
\label{eq10}
\end{equation}
we have the following approximate expression for the energy levels of the quantum bouncer with quadratic dissipation
\begin{equation}
E_{n}=mg\ell_{g}z_{n}-\frac{16}{15}\gamma g\ell_{g}^2(1-e^{-2\gamma d/m})z_{n}^2,
\label{eq11}
\end{equation}
where $\delta E_{n}^{(1)}<0$ as one would expect from the dissipation in the system. In Fig. (\ref{Fig1}) we show the graph for the energy loss for a neutron in a gravitational field. From the figure one can see that Eq. (\ref{eq11}) is only valid for the first quantum states. 
\\
\section{Conclusions}
\label{sec:Con}
We have shown that for the particle in a uniform gravitational field and with a dissipative force proportional to $v^2$ we can construct an effective Hamiltonian which corresponds to the energy of the system. We found that it is possible to obtain the energy loss during a full cycle of motion for the quantum bouncer with quadratic dissipation by means of canonical quantization and quantum perturbation theory.
\vspace{-0.5cm}



\begin{thebibliography}{11}

\bibitem{obs} Valery V. Nesvizhevsky, Hans G. B\"orner, Alexander K. Petukhov, Hartmut Abele, Stefan Bae$\beta$ler, Frank J. Reu$\beta$, Thilo St\"oferle, Alexander Westphal, Alexei M. Gagarski, Guennady A. Petrov and Alexander V. Strelkov, {\it Nature}, {\bf 415}, 297-299 (2002).

\bibitem{obs1} A. Westphal, H. Abele, S. Bae$\beta$ler, V. V. Nesvizhevsky, K. V. Protasov, A. Y. Voronin, {\it Eur. Phys. J. C}, {\bf 51}, 367-375 (2007).

\bibitem{obs2} V. V. Nesvizhevsky,  A. K. Petukhov, H. G. B\"orner, T. A. Baranova, A. M. Gagarski, G. A. Petrov, K. V. Protasov, A. Yu. Voronin, S. Bae$\beta$ler, H. Abele, A. Westphal, L. Lucovac, {\it Eur. Phys. J. C}, {\bf 40}, 479-491 (2005).

\bibitem{obs3} C. G. Aminoff, A. M. Steane, P. Bouyer, P. Desbiolles, J. Dalibard, C. Cohen-Tannoudji, {\it Phys. Rev. Lett.}, {\bf 71}, 3083-3086 (1993).

\bibitem{GG} G. Gonz\'alez, {\it Int. J. of Theo. Phys.}, {\bf 43}, 1885-1890 (2004).

\bibitem{GG1} G. Gonz\'alez, {\it Int. J. of Theo. Phys.}, {\bf 46}, 417-423 (2007).

\bibitem{GG2} G. Gonz\'alez, {\it Int. J. of Theo. Phys.}, {\bf 46}, 486-491 (2007).

\bibitem{GG3} G. L\'opez and  Gonz\'alez, {\it \it Int. J. of Theo. Phys.}, {\bf 43}, 1999-2008 (2004).

\bibitem{obs4} F. Negro and A. Tartaglia, {\it Phys. Rev. A}, {\bf 23}, 1591-1593 (1981).

\bibitem{obs5} F. Negro and A. Tartaglia, {\it Phys. Lett. A}, {\bf 77}, 1-2 (1980).

\bibitem{obs6} C. Stuckens and D. H. Kobe, {\it Phys. Rev. A}, {\bf 34}, 3565-3567 (1986).

\bibitem{obs7} J. S. Borges, L. N. Epele, H. Fanchiotti, {\it Phys. Rev. A}, {\bf 38}, 3101-3103 (1988).

\bibitem{obs8} M. Razavy, {\it Phys. Rev. A}, {\bf 36}, 482-486 (1987).

\bibitem{obs9} Dharmesh Jain, A. Das, Sayan Kar, {\it Am. J. Phys.}, {\bf 75}, 259-267 (2007).

\bibitem{obs10} D. M. Goodmanson, {\it Am. J. Phys.}, {\bf 68}, 866-868 (2000).

\end{thebibliography}
\end{document}